\documentstyle[12pt]{article}

\title{On the Pauli principle violation in QFT}
\author{Victor Novikov \\
ITEP, Moscow, Russia}
\date{}

\def\fun#1#2{\lower3.6pt\vbox{\baselineskip0pt\lineskip.9pt
\ialign{$\mathsurround=0pt#1\hfil##\hfil$\crcr#2\crcr\sim\crcr}}}

\begin{document}

\maketitle

\begin{abstract}

We propose a new mechanism for a  ''small" violation of Pauli Principle in the framework of Quantum Field Theory.
Instead of modification of algebra -  commutation relations for fields - we introduce spontaneous violation
of Pauli Principle which is proportional to the vacuum fermionic condensate.

\end{abstract}

\newpage

Getting older some of the theorists  turn to  the foundations of
Quantum Mechanics. This is exactly my case. I am going to discuss
the possibility of  Pauli Principle breaking - one of the
cornerstone of Quantum Field Theory (QFT). My talk is based on
yet unpublished paper (un)written  in collaboration with my old
friend Sasha Dolgov (at early stage  Maxim Pospelov participated
to our discussions).

\section{Introduction}

The great merit of Pauli Principle is that it can be formulated in terms  understandable to any person from the street.
On the other hand  the proof of  Pauli Principle is based on a rather advanced formalism of QFT  comprehensible
to the tiny fraction of experts.  Feynman wrote in his famous
Lectures on Physics:``It appears to be one of the few places in physics where there is a rule which can be stated very simply,
but for which no one  has found a simple and easy explanation. The explanation is deep down in relativistic quantum mechanics.
This probably means that we do not have a complete understanding of the fundamental principle involved." \cite {Feynman}.

To understand principle better sometime it is  useful to  break it.
I got involved in this business when I have seen the paper by Dolgov and Smirnov \cite{Dolgov}, where  the authors
proposed a fractional statistic for neutrino. (Fractional statistics for charged particles such as electron is excluded by
experiment). In this way they wanted to get a Bose  condensation of neutrino in vacuum and  explain the origin of dark matter.
My recollection was that it is absolutely impossible to get Pauli Principle violation in the framework of QFT and
that  it is well known in literature that it is impossible. I was wrong - it is possible and there is vast literature on this subject.

\section{Short History of Discovery and Long History of Breaking}

Exclusion Principle was introduced by Pauli in 1925 \cite {Pauli1}.  The first formal proof in QFT was developed by him fifteen years later \cite {Pauli2}.
The long list of papers that improved and purified the original proof can be found in any book on axiomatic field theory. The best  reference, as I know,
is still the book of  R.F.Streater  and  A.S.Wightman \cite {SW}.

Non-standard types of statistics such as parastatistics had been known for a long time \cite{Greenberg}.
It was so to say "large"  violation of Pauli Principle.
Near a dozen or papers were published on this subject till 1987.
In 1987 the first model for \underline{small}  violation of Pauli Principle
was constructed in the framework of QM \cite{Kuzmin1}.  In this model the bilinear commutation relations for annihilation and creation operators were
modified to trilinear relations with small parameter. It was a great success. But  it was found immediately that generalization of bilinear relations to trilinear ones
is rather difficult procedure in QFT . This modification
inevitably leads to some pathology  and consistent QFT with fractional statistics does not exist \cite{Okun}.
After this work near a hundred of papers with different modifications of Algebra of Operators were published.
In a very recent paper \cite{Kuzmin2} a kind of  "No-Go Theorem"  was proven. It was found that it is difficult (if possible) to get  a \underline{small}
violaion of Pauli Principle in QFT with bilinear algebra. \footnote{This paper contains rather complete list of references on the Pauli Principle Violation.}

We suggest not to modify Algebra of Fields and not to destroy  Pauli Principle by brute force.
 Instead of genuine breaking we suggest to imitate Pauli Principle violation  exactly like
"Spontaneous Symmetry Breaking" imitates symmetry violation.

\section{Simulation of Symmetry Breaking}

The phenomenon of  Spontaneous Symmetry Breaking is getting known  to students from a course on general physics
when they learn about ferromagnetic material. Suppose that we have a piece of iron. Electromagnetic interaction of
electrons  in metal is  $O(3)$ invariant. Suppose now that we switch-on an external   electric current $\bf {j}$.
Electric current  produces magnetic field $\bf B$. It is clear that  interaction of non-zero external current  with electrons
in metal breaks $O(3)$ invariance.
If we switch-off the current the $O(3)$ symmetry will be restored. On the other hand due to magnetization of ferromagnet
the external magnetic field $\bf B$ still remains non-zero.   As a result we get  $O(3)$ invariant system that interacts with $O(3)$ non-invariant
external field. This effect has a name "Spontaneous Symmetry Breaking" though symmetry is not broken. We get
a sort of imitation  of  $O(3)$ symmetry violation.

The  standard description of Spontaneous Symmetry Breaking in QFT step by step follows that example.
First one considers a system with  internal symmetry $G$ and Lagrangian $L_{SYM}(\phi)$,
where $\phi$ is a general symbol for fields. We suppose that  fields  $\phi$ are transformed in some nontrivial way under
symmetry transformation $G$.
At second step one switches-on an external classical source (current)  $J$ for field. In this case the total Lagrangian is
\begin{equation}
L=L_{SYM} + J \phi,
\end{equation}
and external source breaks the symmetry.
If  the current  is nonzero, i.e. $<J>\not=0$, it produces  nonzero classical field in vacuum, i.e.
\begin{equation}
<\phi>_{J} =\phi_{cl}
\end{equation}
Fluctuations  near this classical fields are described by $\phi_{qu}$

\begin{equation}
\phi_{qu}=\phi -\phi_{cl}.
\end{equation}
They interact with non-invariant object - external classical field.

At last step one switches-off the current, i.e. one puts $J\equiv 0$. In this case the term that breaks the symmetry
goes to zero $J\phi=0$.
It happens that  for some systems  the equation of motion for  v.e.v. of the field $\phi$ is non-zero even for zero current $J=0$, i.e.
\begin{equation}
<\phi>_{0} =\phi_{cl}\not=0.
\end{equation}
That is non-zero vacuum condensate.

Symmetric interaction of fluctuations with non-symmetric condensate
produces imitation of symmetry breaking. This is Nambu-Goldstone mechanism of spontaneous symmetry breaking.

\subsection {Scalar Condensate}

In QFT  there exists a unique explicite example of self-interaction that produces  a condensation of scalar fields.
It is enough to take a special potential for scalars

\begin{equation}
V(\phi) =\lambda [\phi^2 - \phi_{cl}^{2}]^2.
\end{equation}

In the Standard model we use similar  potential for Higgs fields to produce vacuum condensate.
As a result propagating $SU(2)$ and $U(1)$ massless gauge bosons and massless Goldstone bosons
scatter on this  vacuum condensate. There is also inelastic scattering on the condensate
that mixes  vector gauge bosons with
Goldstone scalar bosons.  The net effect is that instead of massless gauge and Goldstone bosons one gets massive
gauge bosons
\subsection{Vector Condensate}

There exist a vast literature on a Lorentz symmetry breaking. Fortunately in the case of $QED$ any "reasonable" violation of Lorentz symmetry
leads to a very simple modification of the Standard Lagrangian $L$:

\begin{equation}
\delta{L} = g \epsilon _{\mu\nu\alpha\beta}n_{\mu} A_{\nu} F_{\alpha\beta}
\end{equation}
where $ A_{\nu}$ is four-potential for e.m. field, $  F_{\alpha\beta}$ is e.m. field-strength tensor, and
numerical vector  $n_{\mu}$ breaks Lorentz symmetry. There are two schools of thinking: one treats  $n_{\mu}$
as an external vector, and the second one treats it as a vacuum
condensate of some vector field  $ B_{\mu}$,  i.e.  $n_{\mu}\equiv< B_{\mu}>_0$.
The explicite mechanism for vector field condensation  in vacuum  does not exist   in 4D QFT.
Nowadays it is not a great disaster.  One can consider our space-time as a 4D brane  in multi-dimensional world
Vector field $ B_{\mu}$ can be a zero mode living on this brane.

In any case  light propagates through vector condensate and has different refraction indexes for left- and right-polarized photons.
That is an explicite violation of Lorentz symmetry and $CPT$ symmetry.

\subsection{Fermion condensate}

Let we introduce a source for fermions and change an initial Lagrangian:

\begin{equation}
L \longrightarrow L + \bar J\psi  + \bar \psi J,
\end{equation}
where $J, \bar J$ are ''classical" currents for fermions,  i.e. some grassmanian numbers.  The introduction of a source term for fermions is rather
standard trick in QFT.  In this way one can constract partition  function $Z(J,\bar J)$  and generate  the complete set of
fermionic correlators  $<\psi_{1}... \bar \psi_{n}>$
as a variational derivatives of  $Z(J,\bar J)$ over  currents  $J$,$ \bar J$  at $J = 0$.

In the case of  $J \not= 0$  the  nonzero current generates nonzero expectation value of
the fermionic field:
\begin{equation}
<\psi>_{J} = \xi \not= 0,
\end{equation}
where $\xi$ is also a grassmanian number.  Nonzero  value of $\xi$ violates Lorentz and rotational symmetry. It is interesting to understand whether it possible to have nonzero
value of  $\xi$ at zero current $J=0$. Actual mechanism for spontaneous breaking of symmetry, i.e. for vacuum condensation of $\xi$,  is unknown.
In the standard QFT in four dimension it is impossible. But we can think about our space-time as a four-dimensional brane in multi-dimensional space
with fermionic zero mode living on the brane.
Another possibility is that our Lord just forgot to switch-off the fermionic current $J$.

In any case we will assume that $\xi$ is nonzero. Certainly it violates  Lorentz symmetry.
In addition it should violate the Pauli Principle.
Indeed consider a simple QFT model.
\begin{equation}
L=\bar \psi(\hat{p}-m)\psi +1/2 \phi(\hat{p}^2-m^2)\phi + \lambda \phi (\bar \psi \psi) + \bar J\psi  + \bar \psi J
\end{equation}
where $\phi(x)$ and $\psi(x)$ are neutral boson and fermion  fields.
Equations of motion looks like
\begin{equation}
(\hat{p}-m)\psi + \lambda \phi  \psi + J = 0; \;\;\;
(\hat{p}^2-m^2)\phi + \lambda  (\bar \psi \psi) =0.
\end{equation}
For classical nonzero constant current

\begin{equation}
J(x) \equiv  J = m \xi \not= 0
\end{equation}
we get that
\begin{equation}
<\psi>_{J} \equiv \xi,
\end{equation}
\begin{equation}
<\phi>_{J} \equiv \frac {\lambda}{m^2}\bar \xi\xi
\end{equation}
The propagation of the excitations in the vacuum with these two condensates
\begin{equation}
\psi = \xi + \psi_q ; \;\;\;  \phi = \frac {\lambda}{m^2}\bar \xi\xi +  \phi_q,
\end{equation}
is described by the quadratic form
\begin{equation}
L^{(2)}=\bar \psi_q(\hat{p}-\bar m)\psi _q+\frac{1}{2} \phi_q(\hat{p}^2-m^2)\phi_q + \lambda \phi_q [\bar \xi \psi_q + \bar \psi_q  \xi],
\end{equation}
where $\bar m = m - \frac {\lambda^2}{m^2} \bar\xi\xi$.  It is clear that the last term proportional to $\lambda$
describes inelastic scattering on the fermionic condensate that transforms fermions into bosons and visa verse:
\begin{equation}
Bosons  \Longleftrightarrow Fermions
\end{equation}

Evidently such transformation breaks statistic. The direct way to calculate statistic of the excitations that corresponds to $\psi_q$ and $\phi_q$
is to diagonalize this quadratic form. Technically this rather tricky problem.
The proper way to diagonalize it is to quantize a system in a box. In this way  we reduce QFT problem to QM problem of the one given level.

\section{QM model}
 Consider field operators $\phi(x)$ and $\psi(x)$ as an expansion over  plane waves in the box

\begin{equation}
\phi(x) = \sum_{p} \frac{1}{\sqrt{2\omega(p)}} [ a(p)  exp {(i px)} + a^{+} (p) exp {(-i px)}],
\end{equation}
and
\begin{equation}
\psi(x) = \sum_{p}  \frac{1}{\sqrt{2\omega(p)}} [ b(p) u(p) exp {(i px)} + h.c.],
\end{equation}
where $\omega^2={\bf p}^2 + m^2$, and $(a(p), a^{+} (p))$ and  $(b(p) , b^{+} (p))$ are annihilation and creation operators for the original
scalar and spinor fields.  For the mode with given 3-momenta $\bf p$ we have a system with two degrees of freedom, i.e. simple Quantum Mechanics
\begin{equation}
H = \omega(p) [ a a^{+} + b b^{+}] + \lambda [ a^{+}\zeta^{+} b + b^{+} \zeta a],
\end{equation}
with grasmanian parameter $\zeta=(\bar u \xi)/2\omega$ and creation and annihilation operators that satisfy well-known algebra
\begin{equation}
[ a , a^{+}]_{-}= [ b, b^{+}]_{+} =1; \;\;\; [ a , a]_{-}= [ b, b]_{+} =[ a, b]_{-} =[ a, b^{+}]_{-}=0.
\end{equation}
One can verify that this algebra is invariant under one-parameter group of invariance $(a, b) \rightarrow (A,B)$:
\begin{equation}
a=[1-\frac{1}{2}(\beta^{*}\beta)(\zeta^{*}\zeta)]A + \beta (\zeta^{*}B)
\end{equation}
\begin{equation}
b= -\beta^{*} A\zeta +[1+\frac{1}{2}(\beta^{*}\beta)(\zeta^{*}\zeta)] B,
\end{equation}
where $\beta$ is an arbitrary complex number.

Operators $A, B$ satisfy the same algebra
\begin{equation}
[ A , A^{+}]_{-}= [ B, B^{+}]_{+} =1; \;\;\; [ A , A]_{-}= [ B, B]_{+} =[ A, B]_{-} =[ A, B^{+}]_{-}=0.
\end{equation}

To diagonalize quadratic Hamiltonian we should take
 \begin{equation}
\beta=\beta^{*}= - \lambda/2\omega(p).
\end{equation}
In the new canonical coordinates Hamiltonian looks like a sum of bosonic oscillator and fermionic oscillator:
\begin{equation}
H = \omega_{1} A A^{+} + \omega _{2}B B^{+}
\end{equation}
with
\begin{equation}
 \omega_{1} = \omega + \frac{\lambda^{2} m }{16 \omega^3} \bar \xi \xi,
\end{equation}
\begin{equation}
 \omega_{2} = \omega -\frac{ \lambda^{2 }m}{16 \omega^3} \bar \xi\xi.
\end{equation}
Mixing between bosonic  subsystem $(a,a^{+})$ and fermionic subsystem $(b, b^{+})$
 leads to a repulsion of the levels (with center of mass constrain)
 \begin{equation}
 \omega_{1}+\omega_{2}= 2 \omega.
\end{equation}
\section{Statistics}
In terms of diagonal variables the spectrum of Hamiltonian is known and one can  calculate the average number of  particle
at given state using the standard rules of Statistical Mechanics.
For particles that are created by operator $A^{+}$ we get a canonical Bose distribution:
\begin{equation}
 <N>_{Bose} = < A A^{+}> = \frac{1}{exp(\omega_{1}/T) - 1}.
\end{equation}
 with shifted frequency $\omega_1$.
For particles that are created by operator $B^{+}$ we get a canonical Fermi distribution
\begin{equation}
 <N>_{Fermi} = < B B^{+}> = \frac{1}{exp((\omega_{2}-\mu)/T) + 1},
\end{equation}
where $\mu$ is a chemical potential.
These are the distributions for the  diagonal states.

In terms of the  initial particles that are created in collisions the same equations look like a mixed statistic .
 Indeed if we introduce distributing numbers for initial particles
\begin{equation}
 <n>_{B} = < a a^{+}>,
\end{equation}
and
 \begin{equation}
 <n>_{F} = < b b^{+}>,
\end{equation}
we get
\begin{equation}
 <n>_{F} = (1+ \beta^2 \zeta^{+}\zeta) N_{Fermi} - \beta^2 \zeta^{+}\zeta N_{Bose},
\end{equation}
and
\begin{equation}
 <n>_{B} = (1- \beta^2 \zeta^{+}\zeta) N_{Bose} + \beta^2 \zeta^{+}\zeta N_{Fermi},
\end{equation}
where $ \beta = -\lambda/2\omega$,

As a result for the distribution of initial "neutrinos"  we get
\begin{equation}
 <n>_{\nu} = [1+ O(\lambda\bar \xi \xi)] \frac{1}{exp((\omega-\mu)/T) + 1} - \frac {\lambda^2 m }{32\omega^4} \bar \xi \xi  \frac{1}{exp(\omega/T) - 1},
\end{equation}
i.e. a piece of a standard Fermi distribution with slightly modified frequency plus a fraction of Bose distribution. The admixture of Bose statistic is proportional to
the condensation of Fermi field $\bar \xi \xi$ and can be made arbitrary small.

\section {Back to QFT}
 In terms of field variables
\begin{equation}
\phi(x) = \sum_{p} \frac{1}{\sqrt{2\omega(p)}} [ a(p)  exp {(i px)} + a^{+} (p) exp {(-i px)}],
\end{equation}

\begin{equation}
\psi(x) = \sum_{p}  \frac{1}{\sqrt{2\omega(p)}} [ b(p) u(p) exp {(i px)} + h.c.]
\end{equation}
 the transformation  of operators $(a,b)\Rightarrow (A,B)$ looks like non-local transformations:

\begin{equation}
\phi(x) \Rightarrow [1-\frac{\lambda^{2}m}{64}\bar \xi \xi \frac{1}{(-\nabla^2+m^2)^2}] \phi(x)-\frac{\lambda}{4} \frac{1}{(-\nabla^2+m^2)}[\xi\bar\psi+\psi\bar\xi]
\end{equation}
\begin{equation}
\psi(x) \Rightarrow [1+\frac{\lambda^{2}m}{64}\bar \xi \xi \frac{1}{(-\nabla^2+m^2)^2}] \psi(x)+\frac{\lambda m}{4}  \frac{1}{(-\nabla^2+m^2)}\phi(x)\xi.
\end{equation}
By construction it is clear that these non-local transformations do not violate causality.

\section{Conclusions}

Let me summarize the results.

The Fermion vacuum  condensate simulates Pauli Principle breaking. This is a new way to play with Pauli Principle Breaking in QFT.
 We hope that we have made a step in true direction.

\section {Acknowledgments}
 I would like to thank  the organizers of La Thuile conference, particular  Mario Greco,
 for their warm hospitality and for excellent conference.

This research was partly supported by RFBR grant 07-02-00021

\end{document}